\documentstyle[aps,prl,twocolumn]{revtex}
\input epsf
\input rotate
\input psfig.sty

\begin{document}
\draft
\title{A continuum model for low temperature relaxation of crystal steps
}
\author{O.\ Pierre-Louis}
\address{Laboratoire de Spectrom\'etrie Physique-Grephe, CNRS, UJF-Grenoble 1,
BP87, F38402 Saint Martin d'H\`eres, France}
\date{\today}

\maketitle
\begin{abstract}
High and low temperature relaxation
of crystal steps are described in a unified picture,
using a continuum model based on a modified expression of 
the step free energy. Results are
in agreement with experiments and Monte Carlo
simulations of step fluctuations and monolayer cluster
diffusion and relaxation. In an extended model
where mass exchange with neighboring terraces is allowed,
step transparency and a low temperature regime for 
unstable step meandering are found.
\end{abstract}
\vspace{0.3in}
\pacs{PACS numbers: 05.40.-a, 61.46.+w, 68.35.Fx, 36.40.Sx}
\vspace{-0.3cm}

The upsurge of nanotechnologies and
of microscopic visualization techniques (such as
Scanning Tunelling Microscopy) in the past
15 years, raises the challenge for
modeling crystal surfaces morphology 
at smaller and smaller scales.
Surface dynamics crucially depends on the
underlying microscopic discreteness,
especially in non-equilibrium or low temperature conditions.
Accounting for the presence of crystal steps has allowed
some important breakthroughs in the description
of growth and relaxation of nanostructures \cite{bcf,vill}.
In this Letter, it is pointed out that steps
themselves exhibit a discrete sub-structure
that drastically affects their dynamics at low temperatures,
and that a continuum model based on a 
modified free energy allows one to account 
for this low temperature regime.

An isolated step is always rough: steps do not have macroscopic
facets or sharp angles (see Ref.\cite{vill} for a discussion).
This statement has
motivated a modeling for mass transport along steps (edge diffusion), 
first developed by Mullins \cite{mullins},
based on a description
at lengthscales larger than the distance between kinks.
Measurement of the time correlations of 
fluctuating steps \cite{giesen} has revealed a
low temperature regime for edge diffusion not explained by this model.
Systematic low temperature deviations from its predictions, 
have also been reported in experimental
and kinetic Monte Carlo (MC) studies  
of monolayer island diffusion \cite{metiu,pai},
and relaxation from a deformed shape \cite{jensen}.

In the following, a continuum model is presented,
based on a modified expression of the step free energy
that explicitly accounts for edge-atoms ({\it i.e}
mobile atoms at the steps).
This model exhibits  high and low temperature regimes
in agreement with kinetic MC simulations and experiments.
When mass exchange with neighboring terraces is allowed,
it also accounts for the recently observed 
step transparency (also called permeability) 
on high temperature Si(111) surfaces \cite{metois},
and for low temperature wavelength selection  
of unstable step meandering during growth \cite{ernst}.

The free energy of a step is traditionally taken to be
proportional to its length \cite{vill}:
\begin{eqnarray}
{\cal F}_0=\int dx \gamma (1+(\partial_x\zeta)^2))^{1/2}
\label{e:free_en_M}
\end{eqnarray}
where $\zeta(x,t)$ is the meander of the step
with respect to the straight step configuration,
and $\partial_x$ denotes the partial derivation with respect to $x$.
At low temperatures, one can neglect atom
detachment from steps to terraces \cite{giesen}.
Mass transport then only occurs via mobile edge-atom diffusion
along the step edge, and is driven by gradients of the chemical potential
$\mu=\Omega(\delta {\cal F}_0/\delta \zeta)=\Omega\tilde{\gamma}\kappa$
where $\Omega$ is the atomic area, $\delta/\delta\zeta$ denotes 
a functional derivative with respect to $\zeta$, 
and $\tilde{\gamma}=\gamma+\gamma''$
is the step stiffness.
Step motion results from the divergence of the local
mass flux $j=-aD_L\partial_x\mu/k_BT$, where $D_L$ is the macroscopic
diffusion constant for mass transport along the step. Thus,
\begin{eqnarray}
\partial_t \zeta  = \partial_x\left[{aD_L \over k_BT}\partial_x
\left(\Omega \tilde{\gamma} \kappa \right)
\right] \, .
\label{e:M}
\end{eqnarray}
Eq.(\ref{e:M}) is the usual starting point for studies of
edge diffusion-driven step dynamics. It is always valid 
for lengthscales larger than the distance between kinks,
and long enough timescales.

As an example, let us consider a [110] step on a Cu(100) surface,
where the kink energy is $E_k=0.13$ eV \cite{giesen}. At low temperatures,
kink density is $N_k\approx 2{\rm exp}(-E_k/k_BT)/a$,
where $a$ is the lattice spacing.
The distance between kinks is then $N_k^{-1}\approx 75a$ for $T=300$K, and
$N_k^{-1}\approx 10^3a$ at $T=200$K.  Hence, not only $N_k^{-1}\gg a$,
but at low enough $T$, $N_k^{-1}$ may be much larger
than observation lengthscales:
Scanning Tunelling Microscopy nowadays allows one to
study step fluctuations up to atomic scales \cite{giesen,frenken}, and
monolayer clusters of several nanometers are 
observed \cite{pai}. Moreover, step relaxation timescales related to
edge-atom motion from kink to kink can also become large
when $N_k$ is small.

The very different role played by mobile edge-atoms and
atoms incorporated into the solid suggests that
the step should rather be described as a heterogenous phase
at scales smaller than $N_k^{-1}$.
Hence, the total free energy shall be written as:
\begin{eqnarray}
{\cal F}=\int dx \left[ \gamma (1+(\partial_x\zeta)^2))^{1/2}
+{\alpha \over 2} (c-c_{eq}^0)^2 \right]
\label{e:free_en}
\end{eqnarray}
for a step wandering about the closed-packed orientation $\hat{x}$.
$c(x,t)$ is the macroscopic concentration of mobile edge-atoms
along the step, as presented in Fig.1.
The  term $(c-c_{eq}^0)^2$ accounts for
the departure from local equilibrium, 
and will be seen to be irrelevant 
at high temperature.
We shall see in the following how step relaxation based on
Eq.(\ref{e:free_en}) accounts for high and low $T$
regimes in a wide variety of physical situations.
Kinks are not explicitly described in this 
model. Nevertheless, the relaxation of $c$ to its
equilibrium value $c_{eq}^0$ implicitly involves
the kink distribution, as will be seen later.
\begin{figure}
\centerline{
\hbox{\psfig{figure=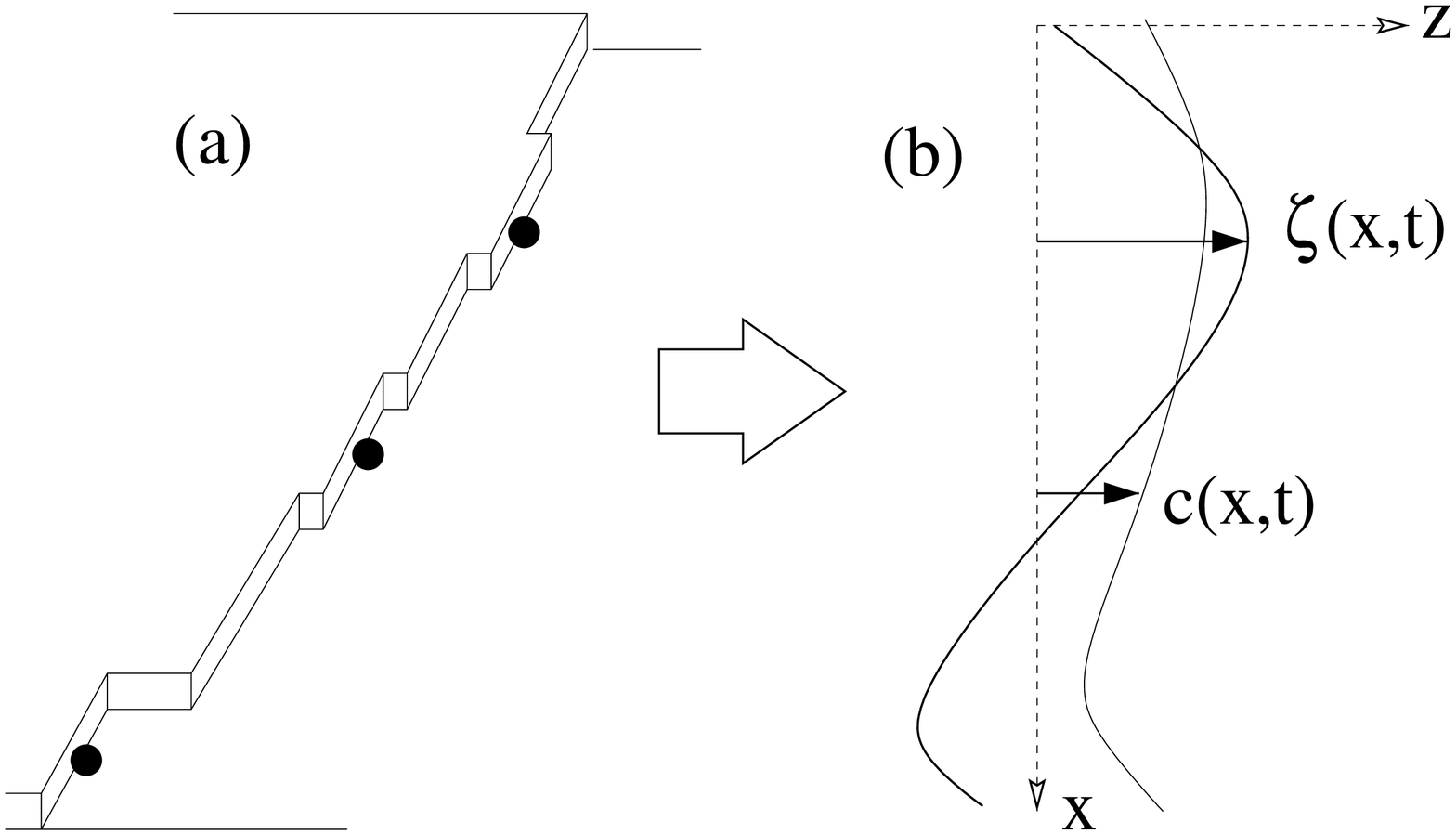,width=6 cm,angle=0}}}
   \caption{
The step position $\zeta(x,t)$ and the macroscopic edge-atom concentration
$c(x,t)$ are both needed in a continuum description
for low temperature dynamics of crystal steps.}
\label{fig1}
\end{figure}      
For small perturbations, Eq.(\ref{e:free_en}) is expanded
to second order in $\zeta$ and $u=c-c_{eq}^0$. At thermal equilibrium,
equipartition implies that each Fourier mode $u_k$ or $\zeta_k$ 
of wavevector $k$ carries the same amount of energy:
\begin{eqnarray}
{\tilde{\gamma} \over 2}k^2\langle |\zeta_k|^2\rangle
={\alpha \over 2}\langle |u_k|^2 \rangle
={k_BT \over 2} \, .
\label{e:stat_spec}
\end{eqnarray}
As a first mean field approximation, edge-atoms can be considered
as non-interacting; thus \cite{landau} $\langle|u_k|^2\rangle=c_{eq}^0$,
and $\alpha=k_BT/c_{eq}^0$.

Let us now turn to the dynamics.
In a way similar to model C in
critical phenomena \cite{hohen}, two evolution equations are written.
The first one for the non-conserved step position $\zeta$ reads:
\begin{eqnarray}
{\partial_t \zeta \over \Omega} 
&=& -A {\delta {\cal F} \over \delta \zeta} +\eta
\label{e:h_nonc}
\end{eqnarray}
where $A$ is a kinetic coefficient, and $\eta$ is a Langevin force.
A second evolution equation is written 
for the conserved total concentration of
atoms ({\it i.e.} atoms in the solid and edge-atoms)
$\psi=\zeta/\Omega+c$:
\begin{eqnarray}
\partial_t \psi &=& \partial_x\left[ B 
\partial_x \left( {\delta {\cal F} \over \delta \psi} \right)
-q \right] 
\label{e:psi_cons}
\end{eqnarray}
where $B$ is a mobility, and $q$ is a conserved noise.
Using Eq.\ (\ref{e:free_en}) in Eqs.\ (\ref{e:h_nonc}-\ref{e:psi_cons}), 
a set of coupled evolution equations is found:
\begin{eqnarray}
{\partial_t \zeta\over \Omega} &=& \nu (c-c_{eq})+\eta
\label{e:motion_straight1}
\\
\partial_t c &=& \partial_x\left[ B \partial_xc-q \right]
-\nu (c-c_{eq})-\eta
\label{e:motion_straight2}
\end{eqnarray}
where $\nu=A \alpha$ , and $c_{eq}$ is found to obey a Gibbs-Thomson
relation:
\begin{eqnarray}
c_{eq}=c_{eq}^0 (1+ \Gamma \kappa) \, ,
\label{e:GT_line}
\end{eqnarray}
where $\kappa$ is the step curvature, and
$\Gamma=\Omega \tilde{\gamma}/k_BT$
is the capillary length.
The obtained equations share similarities 
with the model of Appendix A of Ref.\cite{khare2}
for line diffusion or that of Ref.\cite{cahn} for
surface diffusion.
Nevertheless, they were not derived from
a free energy as done in this Letter, and to our knowledge,
the results mentioned in the following
have not been pointed out by these authors.
Following  Ref.\onlinecite{oplcm},
the correlations of the Langevin forces are found
within a local thermodynamic equilibrium approximation:
\begin{eqnarray}
\langle \eta(x,t)\eta(x',t') \rangle &=& 2\nu c(x,t)\,
\delta(x-x') \delta(t-t') \, ,
\nonumber \\
\langle q(x,t) q(x',t') \rangle &=& 2B c(x,t)\,
\delta(x-x') \delta(t-t') \, .
\label{e:correl}
\end{eqnarray}

As a first approach, we now propose some phenomenological
expressions for the kinetic coefficients.
The macroscopic diffusion constant of edge-atoms along the steps
is approximated by that of a tracer edge-atom on a frozen
step with a given kink density $N_k$. 
Defining the diffusion constant of mobile edge-atoms between
kinks $D$ and the kinetic attachment lengths 
$d_\pm=a({\rm exp}(E_\pm/k_BT)-1)$, where $E_\pm$ are the additional
energy barrier (with respect to diffusion) 
for atoms to stick to a kink from both sides,
and using the result of Ref. \onlinecite{natori}
it is found that:
\begin{eqnarray}
B= {D \over 1 + N_k(d_++d_-) } \, .
\label{e:macro_diff}
\end{eqnarray}
The macroscopic attachment coefficient $\nu$ is the inverse of 
the relaxation time of the concentration,
which is related to the timescale of diffusion of a mobile atom from
one kink to another. Hence,
\begin{eqnarray}
\nu \approx BN_k^2 \, .
\label{e:macro_nu}
\end{eqnarray}

We shall first address the case of a straight step
at equilibrium fluctuating about the close-packed direction $x$.
The time correlation function
\begin{eqnarray}
G(t)=\langle [\zeta(x,t)-\zeta(x,t+\tau)]^2 \rangle
\label{e:time_corr}
\end{eqnarray}
has been measured in experiments and MC simulations \cite{giesen}.
This quantity is evaluated within
the quasistatic approximation, which stipulates that
the edge-atom concentration reaches a steady state 
on time scales much shorter than kink motion.
Thus the l.h.s. of Eq.(\ref{e:motion_straight2}) vanishes.
Comparing the quasistatic and full dispersion relations
as in Ref. \onlinecite{opltle}, and using Eq.(\ref{e:macro_nu}),
the quasistatic approximation is found to be valid when
\begin{eqnarray}
\Gamma a^2 c_{eq}^0  \ll N_k^{-2} + k^{-2} \, .
\label{e:quasi}
\end{eqnarray}
At low $T$, in a simple bond-counting model
and for a step along a high symmetry orientation,
$\Gamma a^2c_{eq}\sim a^2{\rm exp}(-E_k/k_BT)\ll 
N_k^{-2}\sim a^2{\rm exp}(2E_k/k_BT)$, where $E_k$ is the
kink energy. At high $T$,
$c_{eq}^0\sim 1/a$, $N_k\sim a^{-1}$ 
and $\Gamma \rightarrow 0$.
Hence, the quasistatic limit is valid at all $T$.

Linearizing Eqs.(\ref{e:motion_straight1},\ref{e:motion_straight2}),
in the quasistatic limit, $G(t)$ is easily evaluated.
For large observation timescales, long wavelength fluctuations 
$\lambda \gg N_k^{-1}$ dominate, and
\begin{eqnarray}
G_{long}(\tau)={a^2 \Gamma(3/4) \over \pi} 
(b^2)^{3/4} (Bc_{eq}^0)^{1/4} \tau^{1/4} \, ,
\label{e:time_corr_long}
\end{eqnarray}
where $b^2=ak_BT/\tilde{\gamma}$ is the step diffusivity.
This expression corresponds to the one
given in Ref. \cite{bartelt-masson} starting from the Mullins model
Eq.(\ref{e:M}), with $D_L=aBc_{eq}^0$ as expected from Ref.\cite{mullins}.
For short observation timescales, 
only short wavelengths ($\lambda \ll N_k^{-1}$) 
contribute to $G$, and:
\begin{eqnarray}
G_{short}(\tau)={a^{3/2} \over \sqrt{\pi}}
(\nu c_{eq}^0 b^2)^{1/2} \tau^{1/2} \, .
\label{e:time_corr_short}
\end{eqnarray}
Using Eq.(\ref{e:macro_nu}) and the relation
$b^2\sim N_k$ --valid at low $T$,
the crossover between the two regimes is found to correspond
to $G(t)\sim a^2$.
This result was found by Giesen {\it et al} \cite{giesen},
by means of a discrete random kink model and MC simulations.
From the relation $G(t^*)\sim a^2$,
the cross-over time between the two regimes is
found to be:
\begin{eqnarray}
t^*\sim (N_k^3 a^2 Bc_{eq}^0)^{-1} \, .
\end{eqnarray}
Using numerical values for Cu(11n) vicinal surfaces
given in Ref.\onlinecite{giesen},
one finds $t^*\sim 10^{-19}{\rm exp}(14870/T)$ s, where $T$ is
in Kelvins. With observation times $t^*\sim 1$s \cite{giesen}, the crossover
is found for $T\approx340$K, in quantitative agreement with
experiments \cite{giesen}. Formally, the cross-over from 
high to low temperatures 
(Eqs.(\ref{e:time_corr_long}-\ref{e:time_corr_short}))
looks similar to that from line diffusion to terrace diffusion
reported by several authors \cite{bonzel,khare2}.
Neverteless, it is physically different since there is
only line diffusion here.

Monolayer cluster relaxation and diffusion
might also be addressed by this model.
Let us consider an circular island  
having small perturbations about its mean radius $R_0$ 
defined by  $R(\theta)=R_0+\rho(\theta)$.
The polar coordinates $R$ and $\theta$ are used.
In a 'circular model', $x$ is simply replaced in the model
by $R_0\theta$ for small perturbations.
In the linear approximation, the relaxation time of a small perturbation
$\rho=\epsilon \cos(n\theta)$ is:
\begin{eqnarray}
t_n=2\pi {R_0^2 \over \nu  \Gamma a^2 c_{eq}}
{n^2+R_0^2(\nu/B) \over n^2(n^2-1)} \, .
\end{eqnarray}
A crossover from $t_r\sim R^4_0$
when $R_0\gg N_k^{-1}$ to $t_r\sim R_0^2$ when $R_0\ll N_k^{-1}$
is found in agreement with kinetic MC 
simulations in Ref.\onlinecite{jensen}.

The diffusion constant of a cluster is defined as:
$D_c=\langle {\bf r}_{CM}^2(t) \rangle /4t$,
where ${\bf r}_{CM}$ indicates the position of the center of mass
of the cluster.
The cluster diffusion constant 
is calculated without using the quasistatic limit,
and reads:
\begin{eqnarray}
D_c={a^4c_{eq} \over \pi R_0}
\,{1 \over R_0^2/B+1/\nu}
\label{e:Dc_2f}
\end{eqnarray}
The high $T$ behavior 
$D_c=a^3D_L/\pi R_0^3$, calculated from Eq.(\ref{e:M})
in Ref. \cite{khare,opltle}, is recovered when $R_0\gg N_k^{-1}$,
provided once again that $D_L=aBc_{eq}$. In the low $T$ regime,
where $R_0 \ll N_k^{-1}$, another scaling limit is found:
$D_c \sim R_0^{-1}$, in agreement with previous experimental \cite{pai}
or MC \cite{metiu} studies.
The circular
approximation catches the essential physical point,
which is the existence of orientations for which $N_k^{-1}$
is much larger than the size of the cluster.
Nevertheless, including anisotropy in our model
is needed for a quantitative comparison with low $T$ experiments
and Kinetic MC simulations.

At higher $T$ or during growth, steps exchange mass with terraces.
For low kink concentration $N_k\ll a^{-1}$, 
direct exchange from kink to terrace can be neglected,
and step meandering is weak, {\it i.e.} $\partial_x\zeta\ll 1$.
An extended model may then be written, with Eqs.(\ref{e:motion_straight1}), 
(\ref{e:GT_line}) and:
\begin{eqnarray}
\partial_t c = \partial_x[B\partial_xc]-\nu(c-c_{eq})
+J_++J_-
\label{e:model_exch1}\\
J_\pm =\beta_\pm C_\pm -\nu_\pm c
\label{e:model_exch2}\\
\partial_tC = D_s\nabla^2C+F-C/\tau
\label{e:model_exch3}
\end{eqnarray}
where $C$ is the concentration of adatoms on terraces.
$D_s$ is their diffusion constant, $F$ the incoming flux 
on terraces, and $\tau$ the adatom desorption time.
Langevin forces were omitted 
in Eqs.(\ref{e:model_exch1}-\ref{e:model_exch3}) 
for the sake of clarity in this brief exposition.
$+$ and $-$ designate the lower and the upper
sides of the step respectively,
$\nu_\pm$ and $\beta_\pm$ are kinetic coefficients.
At equilibrium $c=c_{eq}^0$ and $C=C_{eq}^0$, and 
there should be no mass flux (detailed balance)
so that $J_+=J_-=0$.
Thus, $C_{eq}^0/c_{eq}^0=\nu_+/\beta_+=\nu_-/\beta_-$.
Since we address the case of weak meandering, 
exchange mass fluxes between steps
and terraces are given by $J_\pm\approx -D_s\partial_z C_\pm$,
where $z$ is define in Fig.1. 
These latter equations allow to close the model and 
to evaluate the concentration.

In the case of slow diffusion along steps 
({\it i.e.} $B$ small), or for long wavelength
perturbations (larger than $N_k^{-1}$), the first term 
in the r.h.s. of Eq.(\ref{e:model_exch1}) vanishes.
If an additional approximation is made in taking the
variations of the adatom concentration $C$ on terraces
to be much slower than the
relaxation time $(\nu+\nu_++\nu_-)^{-1}$ of the
edge-atom concentration $c$,
we can set $\partial_tc=0$ in Eq.(\ref{e:model_exch1}).
The resulting model may be written:
\begin{eqnarray}
{1\over \Omega}\partial_t \zeta &=&
D_s\partial_zC_+-D_s\partial_zC_-
\label{e:eff_kin_transp1}
\\
D_s\partial_zC_ \pm&=& \pm\tilde{\beta}_\pm(C_\pm-C_{eq})+\beta_0(C_+-C_-)
\label{e:eff_kin_transp2}
\\
C_{eq} &=& C_{eq}^0(1+\Gamma\kappa)
\label{e:eff_kin_transp3}
\end{eqnarray}
where effective kinetic coefficients are defined via
$\tilde{\beta}_+/\beta_+=\tilde{\beta}_-/\beta_-=\nu/(\nu+\nu_++\nu_-)$,
and $\beta_0=\beta_+\nu_-/(\nu+\nu_++\nu_-)$.
With Eq.(\ref{e:model_exch3}) and
Eqs.(\ref{e:eff_kin_transp1}-\ref{e:eff_kin_transp3}), 
we have obtained the standard \cite{metois} model for
``transparent" steps as a special limit. 
Step transparency ({\it i.e} $\beta_0\neq 0$) 
is understood as the possibility
for an atom to attach to a step from a terrace,
and detach to the other terrace
before reaching a kink. Transparency appears as a natural ingredient
of the model equations (\ref{e:motion_straight1},
\ref{e:model_exch1}-\ref{e:model_exch3}).

As a last remark, we shall calculate the most unstable
wavelength for the meandering instability first addressed
by Bales and Zangwill \cite{bz}, with step relaxation 
provided by the full model 
Eqs.(\ref{e:model_exch1}-\ref{e:model_exch3}). 
We use the following parameters:
$\beta_-=0$ and $\nu_-=0$ ({\it i.e.} no mass exchange
with the upper terrace);  $C_{eq}^0=0$, $\nu_+=0$,
and $\beta_+ \rightarrow \infty$ (this implies $C_+=0$).
Moreover, desorption is taken to be vanishingly small: $1/\tau=0$.
The growth rate of a small in-phase perturbation 
$\zeta(x,t)={\rm exp}(i\omega t+ikx)\zeta_{\omega k}$ of all
steps on a vicinal surface reads:
\begin{eqnarray}
i\omega= {
-k^4\Gamma Bc_{eq}^0+F(k\ell{\rm tanh}(k\ell)+{\rm sech}(k\ell)-1)
\over 1+(B/\nu)k^2}
\end{eqnarray}
where $\ell$ is the mean inter-step distance.
The most unstable wavelength is calculated in the long wavelength limit
$k\ell\ll 1$(valid for small fluxes $F$).
One finds: 
\begin{eqnarray}
\lambda_1=4\pi(\Gamma Bc_{eq}^0/F\ell^2)^{1/2}
\end{eqnarray}
when $N_k^{-1}\ll\lambda_1$.
In the opposite case $N_k^{-1}\gg\lambda_1$, a low $T$ 
regime is found, where
\begin{eqnarray}
\lambda_2=2^{1/4}\pi^{1/2}\lambda_1^{1/2}N_k^{-1/2}
\end{eqnarray}
Using activation energies given in Ref.\onlinecite{giesen},
one finds that $\lambda_1$ and $\lambda_2$ follow 
Arrhenius laws with activation energies $0.38$eV 
and $0.12$eV respectively (within $\sim 10$\% error).
The low $T$ regime seems to provide 
the best fit to the experimental result of $0.09$eV \cite{ernst}.
It is not clear though how non-equilibrium line diffusion effects
pointed out in Ref. \onlinecite{oplmrdtle}
combine or compete with these results.

In conclusion, a model has been presented,
based on Eq.(\ref{e:free_en}) and Eq.(\ref{e:macro_nu}),
that accounts both for high and low temperature
step relaxation dynamics observed in
experiments \cite{giesen,pai} and Kinetic MC simulations
\cite{giesen,metiu}.
When mass exchange with neighboring
terraces is added to this model, step transparency appears
as a natural consequence of a low kink density.
We also point out a low temperature regime for step meandering
during growth.

A systematic analysis from a microscopic theory
is still needed for a more rigorous 
evaluation of the kinetic coefficients $B$ and $\nu$.
Moreover, numerical solution of the fully anisotropic
model is needed in order to
describe quantitatively monolayer cluster 
diffusion and relaxation.

Low temperature relaxation of three-dimensional clusters
and nanostructures is a source of long standing controversies
\cite{vill}.
The basic difficulty comes from the singularity
of the free energy for orientations in the vicinity
of a facet. As opposed to this situation,
the free energy of a step does not exhibit singularities.
Thus, a direct generalization of the present study
is not possible. Nevertheless, it provides
some milestones for a continuum description of
three-dimensional clusters
\cite{pimp} and nanostructures relaxation.

Acknowledgments: The author thanks T.L. Einstein for useful
discussions and comments. This work was initiated at the University
of Maryland, College Park, supported by NSF-MRSEC.


\vspace{-0.5 cm}
\begin{thebibliography}{99}
\bibitem{bcf} W.K. Burton, N. Cabrera and F.C. Frank,
Phil. Trans. Roy. Soc. London A {\bf 243}, 299 (1951).
\bibitem{vill} A. Pimpinelli and J. Villain,
{\it Physics of Crystal Growth},
Cambridge Univ Pr, (1999).
\bibitem{mullins} W.W. Mullins, J. Appl. Phys. {\bf 28}, 333 (1957),
and {\bf 30}, 77 (1959).
\bibitem{giesen} M. Giesen-Seibert, F. Schmitz, R. Jentjens, H. Ibach,
Surf. Sci. {\bf 329}, 47 (1995).
\bibitem{metiu} A. Bogicevic, S. Liu, J. Jacobsen, B. Lundqvist,
and H. Metiu, Phys. Rev. B {\bf 57}, R9459 (1998).
\bibitem{pai} W.W. Pai, A.K. Swan, Z. Zhang, and J.F. Wendelken,
Phys. Rev. Lett {\bf 79}, 3210 (1997).
\bibitem{jensen} P. Jensen, N. Combe, H. Larralde, 
J.L. Barrat, C. Misbah, and A. Pimpinelli, 
Eur. Phys. J. B {\bf 11}, 497 (1999).
\bibitem{metois} J.-J. M\'etois and S. Stoyanov,
Surf. Sci. {\bf 440}, 407 (1999).
\bibitem{ernst} T. Maroutian, L. Douillard, and H.-J. Ernst,
Phys. Rev. Lett. {\bf 83}, 4353 (1999),
and private communication.
\bibitem{frenken} L. Kuipers, M.S. Hoogeman, and J.W.M. Frenken,
Phys. Rev. B {\bf 52}, 11387 (1995).
\bibitem{landau} L.D. Landau and M.I. Lifschitz, 
{\it Theoretical Physics V,
Statistical Physics}, Mir Ed., sec.113.
\bibitem{hohen} P.C. Hohenberg and B.I. Halperin, Rev.Mod.Phys.
{\bf 49}, 435 (1986).
\bibitem{khare2} S.V. Khare and T.L. Einstein,
Phys. Rev. B {\bf 57}, 4782 (1998).
\bibitem{cahn} J.W. Cahn and J.E. Taylor, 
Acta Metall.  Mater. {\bf 42}, 1045 (1994).
\bibitem{oplcm} O. Pierre-Louis and C. Misbah,
Phys. Rev. Lett. {\bf 76}, 4761 (1996).
\bibitem{natori} A. Natori and R.W. Godby, 
Phys. Rev. B {\bf 47}, 15816 (1993).
\bibitem{opltle} O. Pierre-Louis, T.L. Einstein,
Phys. Rev. B {\bf 62}, 13697 (2000).
\bibitem{bonzel} Bonzel and Mullins, Surf. Sci. {\bf 350}, 285 (1994).
\bibitem{khare} S.V. Khare, N.C. Bartelt, and T.L. Einstein,
Phys. Rev. Lett. {\bf 75}, 2148 (1995).
\bibitem{bartelt-masson} N.C. Bartelt, J.L. Goldberg, T.L. Einstein,
E.D. Williams, J.C. Heyraud and J.-J. M\'etois,
Phys. Rev. B {\bf 48}, 15453 (1993).
\bibitem{oplmrdtle} O. Pierre-Louis, M.R. D'Orsogna, T.L. Einstein,
Phys. Rev. Lett. {\bf 82}, 366 (1999).
\bibitem{bz} G.S. Bales and A. Zangwill, 
Phys. Rev. B {\bf 41}, 5500 (1990).
\bibitem{pimp} N. Combe, P. Jensen, A. Pimpinelli,
Phys. Rev. Lett. {\bf 85}, 110 (2000).
\end{thebibliography}
\end{document}